\documentclass[tighten, times, twocolumn]{aastex63}
\usepackage{amsmath, graphicx, amssymb, amsfonts, hyperref}
\usepackage[utf8]{inputenc}
\usepackage{textcomp, bm}
\usepackage[toc,page]{appendix}
\graphicspath{{./}{figures/}}

\received{\today}
\revised{\today}
\accepted{\today}
\submitjournal{ApJ}

\shorttitle{RAT alignment of irregular grains}
\shortauthors{Herranen, Lazarian, and Hoang}

\begin{document}
	\title{Alignment of irregular grains by radiative torques: efficiency study}
	
	\author[0000-0001-7732-9363]{Joonas Herranen}
	\affiliation{Department of Physics, University of Helsinki, Gustaf H{\"a}llstr{\"o}min katu 2, 00560 Helsinki, Finland, joonas.herranen@iki.fi}
	
	\author[0000-0002-7336-6674]{A. Lazarian}
	\affiliation{Department of Astronomy, University of Wisconsin, 475 North Charter Street, Madison, WI 53706, USA, lazarian@astro.wisc.edu}
	\affiliation{Center for Computation Astrophysics, Flatiron Institute, 162 5th Ave, New York, NY 10010}
	
	\author[0000-0003-2017-0982]{Thiem Hoang}
	\affiliation{Korea Astronomy and Space Science Institute, Daejeon 34055, South Korea, thiemhoang@kasi.re.kr}
	\affiliation{Korea University of Science and Technology, 217 Gajeong-ro, Yuseong-gu, Daejeon, 34113, South Korea}
	
	\begin{abstract}
		We study the efficiency of grain alignment by radiative torques (RATs) for an ensemble of irregular grains. The grains are modeled as ensembles of oblate and prolate spheroids, deformed as Gaussian random ellipsoids, and their scattering interactions are solved using numerically exact methods. We define the fraction of the grains that both rotate fast and demonstrate perfect alignment with grain long axes perpendicular to the magnetic field. We quantify a factor related to the efficacy of alignment and show that it is related to a $q_{\mathrm{max}}$-factor of analytical model (AMO) of the RAT theory. For the ISRF, our results indicate that the degree of RAT alignment can reach $\sim 0.5$, which may be sufficient to explain observations even if grains do not have magnetic inclusions.
	\end{abstract}
	
	\section{Introduction} \label{sec:intro}
	Measuring the polarization from aligned grains is the most common way of studying magnetic fields in galactic disks and molecular clouds \citep[see][]{Crutcher2010}. Polarization arising from aligned dust interferes with attempts to measure enigmatic B-modes of cosmological origin \citep{Philcox2018}. 
	
	Shortly after the first aligned grains were detected by \citet{Hiltner1949} and \citet{Hall1949}, the first theories of alignment were proposed by \citet{Davis1951} for paramagnetic alignment and \citet{Gold1952b} for mechanical alignment. Those became the two mainstream directions of the research for a couple of decades with paramagnetic alignment being the clear favorite \cite[see][for a historic review]{Lazarian2003b}. Significant contributions to understanding of the complex dynamics of grains in the process of alignment were made by L. Spitzer \citep{Jones1967, Spitzer1979} and E. Purcell \citep{Purcell1969,Purcell1979}. The grains in the studies were assumed to be regular, i.e. spherical or ellipsoidal. 
	
	A very different approach was proposed in \citet{Dolginov1976a} and \citet{Dolginov1976} where it was suggested that an essential piece of physics related to alignment of reallistically irregular grains was missing in the theory. The authors proposed that the irregular grains can demonstrate helicity in the process of their interact with radiation and the interaction can induce their alignment. However, these studies were not able to demonstrate the efficiency of their mechanism and, as it was shown by later research \citep{Hoang2009}, the toy model described by the authors was unable to exhibit  alignment by the radiation. In addition, the theoretical prediction that the direction of grain alignment, i.e. that the grains get aligned parallel or perpendicular to the magnetic field depends on the angle between radiation and the magnetic field was not in agreement with polarization measurements \citep{Andersson2015}. As a result, these pioneering studies were mostly ignored for nearly two decades.
	
	Testing of the idea of the irregular grain interaction with radiation became possible due to the progress in computations as well as due to the development of \texttt{DDSCAT} code by \cite{Draine1994}.  The corresponding paper by \cite{Draine1996a} clearly demonstrated the ability of the radiation to spin up grains to high rotational velocities. For many researchers in the community this already seemed as the solution of the long standing problem of grain alignment. Indeed, one can think that if the grains rotate fast they are difficult to
randomize and the paramagnetic relaxation should perfectly align such suprathermally rotating grains \citep{Purcell1979}. However, the actual situation happened to be more complicated. The study in \citet{Draine1997} showed that in realistic settings the grains experience anisotropic radiation flux which makes the grain dynamics far from trivial. Instead of getting perfectly aligned, grains were shown to exhibit complex phase space dynamics with the effects of paramagnetic relaxation torques being negligible compared to much more powerful torques arising from radiation. The authors did not consider the physics of crossovers \citep{Spitzer1979} and therefore the modeling was not able to reproduce the actual dynamics of grains. These works, nevertheless, established the alignment of grains by radiation as a promising candidate for explaining how grains can be aligned in the interstellar conditions. The actual physics of alignment remained unclear.  
	
	A decade later, after a plethora of studies regarding alignment theory, \citet[henceforth LH07]{Lazarian2007b} provided a physically motivated model of the process which successfully explained the properties of grain alignment that were observed in earlier numerical studies, as well as corrected mistakes and misinterpretations in these studies. While appreciating the complexity the alignment of realistic irregular grains, the authors { identified the RAT alignment of grains with irregular grains being helical} proposed a toy model of a helical grain that allowed analytical description of the alignment. { The corresponding analytic model (AMO) of radiative torques (RAT) alignment was intended to address the mystery why the radiation tends to align grains with long axes perpendicular to magnetic field irrespectively of the direction between the direction of radiation and magnetic field. The latter contradicted to what was suggested in \citet{Dolginov1976a}. In addition LH07 introduced a new type of RAT alignment, i.e. the alignment in respect to the radiation direction rather than to the magnetic field and provided the criterion for this alignment to take place in the intense radiation flows. 
	
	To provide the grain with the property of helicity, LH07 attached at 45 degrees a mirror to an oblate grain and explored the dynamics of this toy system when subjected to the radiation flux.\footnote{This simple model opened a way to predict a new type of alignment, i.e., to predict that the irregular grains can also be aligned with their long axes perpendicular to magnetic field by a gaseous flow and this alignment is similar to the RAT-one (Lazarian \& Hoang 2007b).} The model happened to be {\it surprisingly} successful in reproducing not only general behavior of the irregular grains, but it happened to well reproduce quantitatively the features of the RAT alignment established numerically.} In fact, the predictions of the model were successfully compared with the \texttt{DDSCAT} calculations performed for a {\it limited} sample of irregular grains available for testing. 
	
	Formulation of AMO included identification and analysis of certain key properties of RAT alignment. Specifically, in RAT analysis, the radiative torque is split into three components, out of which the third induces precession of the grain around the radiation field. The ratio of the other two components, $q_\mathrm{max}$, was identified by LH07 to be essential in RAT theory. { The quantity $q_\mathrm{max}$ was found to depend on the shape of the grain, optical wavelength-dependent properties of grain material, and the ratio of radiation wavelength to the effective grain size. }
	
	The subsequent studies made use of the AMO, e.g. \citet{Lazarian2008c, Lazarian2019}, \citet{Hoang2008, Hoang2009, Hoang2016} and they clarified many essential physical processes of the RAT alignment theory. They confirmed the RAT alignment as the dominant process of alignment for grains in various astrophysical environments.  Those include the diffuse interstellar medium (ISM), molecular clouds, photodissociation regions, circumstellar regions, comet comas \citep[see][]{Lazarian2007, Hoang2015, Kolokolova2016, Tazaki2017}. 
	
	With the progress of the predictive RAT theory and its successful testing \citep{Andersson2015} the significance of the RAT mechanism of grain alignment has been generally accepted. { However, the regular deviations from the AMO at different wavelengths were observed within the LH07 study using the limited sample of shapes and wavelengths. To deal with this issue, LH07 proposed the modifications of the AMO, e.g. for UV wavelengths range. More detailed modifications required the studies of more samples, which were not available.} It was also shown in LH07 that the degree of alignment, depend on $q_\mathrm{max}$ which { is a function of several parameters}, the grain shape being the major one. The earlier studies of $q_\mathrm{max}$ were limited to a handful of irregular shapes and it was only in \citet{Herranen2019a} that hundreds of grain shapes were analyzed. The study opened the way for predicting the actual grain alignment for the ensemble of realistic irregular grains, which is what is performed in the present paper.
	
	In Section \ref{sec:formulation}, the grain alignment problem is formulated. In Section \ref{sec:grainmodel}, the grain model used in the numerical methods is introduced. In Section \ref{sec:calculations}, the equations of motion relevant in alignment theory are reviewed. Then, DG and RAT alignment mechanisms are compared using RAT results for ensembles of irregular grains. Finally, the fractions of the ensembles for which high-$J$ attractors exist, $f_{\mathrm{high}J}$, are determined using numerically exact volume-integral-equation-based $T$-matrix method \citep[for further details]{Markkanen2016,Waterman1965}. The implications are discussed in Section \ref{sec:discussion} and results summarized in Section \ref{sec:summary}.
	
	\section{Formulation of the problem}\label{sec:formulation}
	The description of grain alignment can be performed in the phase space as it was first demonstrated in \citet{Draine1997}. The effects of radiative torques on grains is advantageous to analyse in the angular momentum, alignment angle -space, where alignment angle is measured between angular momentum and either the radiation direction or external magnetic field direction.
	In the defined phase space one can find stationary points. Using this approach LH07 provided the study for the AMO and defined, as we discuss below, the parameter space for which the expected grain alignment is perfect, i.e. grain long axes are perpendicular to the magnetic field. Such alignment corresponds to grains in the stable stationary points (attractors) corresponding to high angular momentum, or high-$J$ attractor points. Note, that within the RAT mechanism such alignment is possible without any assistance from the \citet{Davis1951} paramagnetic relaxation. 
	
	If one introduces the measure of alignment of grain angular momentum 
	\begin{equation}
	Q_{J} = \frac{3}{2}\left< \cos^{2}\xi-\frac{1}{3}\right>,
	\label{align_J}
	\end{equation}
	where $\xi$ is the angle between angular momentum and magnetic field direction, then for high-$J$ attractor points $Q_{J, highJ}=1$. For fast rotating grains the internal relaxation aligns the angular momentum ${ J}$ with the axis of maximal moment of inertia ${ X}$ \citep{Purcell1979}. The internal relaxation can arise from the traditional inelastic relaxation \citep{Purcell1979,Lazarian1999b}, or the process termed Barnett relaxation in \citet{Purcell1979} or even more exotic, but even more powerful, nuclear relaxation introduced in \citet{Lazarian1999a}. For the typical interstellar conditions these processes bring ${ J}$ and ${ X}$ in perfect alignment and therefore grains are rotating with their long axes perpendicular to the magnetic field ${ B}$. The corresponding measure of the alignment of grain axes $Q_{X, \mathrm{high}J}$ coincides therefore with $Q_{J, \mathrm{high}J}$, i.e. $Q_{X, \mathrm{high}J}=1$.
	
	For grains in low-$J$ attractor points the alignment of ${ J}$ is not perfect as the subthermal angular momentum can easily be randomized by the gaseous bombardment. In addition, the thermal fluctuations within grains induce the variations of the direction of ${ X}$ in relation to ${ J}$ \citep{Lazarian1994,Lazarian1997b, Lazarian1999a}. Here, we will adopt the value $Q_{X, \mathrm{low}J}\approx 0.25$ that was obtained in numerical simulations in \citet{Hoang2008}.
	
	As a result, the ensemble of grains that has both low-$J$ and high-$J$ attractor points has the total alignment measure (i.e., Rayleigh reduction factor)
	\begin{equation}\label{eq:totaldegree}
	R \approx f_{\mathrm{high}J}Q_{X,\mathrm{high}J} + (1-f_{\mathrm{high}J})Q_{X,\mathrm{low}J}, 
	\end{equation}
	where $f_{\mathrm{high}J}$ is the fraction of grains that are aligned in the high-$J$ attractor points. Above equation assumes that external and internal alignment are independent from each other, and as such is regarded as a first approximation. The overall measure of alignment for an ensemble of grains depends mainly on $f_{\mathrm{high}J}$.
	
	As an important distinction to other current research, Eq. \eqref{eq:totaldegree} implicitly assumes that all grains for which a high-$J$ attractor exists, will be in a high-$J$ state. In steady conditions, collitional excitations will eventually transport grains to a high-$J$ state \citep{Hoang2008, Hoang2016}. However, in recent research \citep{Lazarian2020b}, an additional high-$J$ criterion has been introduced. The so-called $f_{\mathrm{high}J,\,\mathrm{orientation}}$ parameter describes the fraction of grains reaching the high-$J$ state quickly, describing \textit{fast} alignment \cite{Lazarian2007b}. The $f_{\mathrm{high}J,\,\mathrm{orientation}}$ parameter is relevant, e.g. in changing environments, where initially randomly oriented particles may not instantly occupy the high-$J$ state, but require time and interactions to eventually reach the high-$J$ state for grains not in the fraction $f_{\mathrm{high}J,\,\mathrm{orientation}}$, which reach the high-$J$ state \textit{fast}. In the current study, $f_{\mathrm{high}J}$ describes the eventual situation where all grains for which it is possible at all, are in the high-$J$ state.
	
	The existence of high-$J$ attractors depends on the grain interaction with the radiative torques. These torques in LH07 were decomposed into 3 components $\Gamma_1$, $\Gamma_2$ and $\Gamma_3$, where the latter component of the torque is responsible for grain precession, while the two former components are responsible for the alignment. In fact, it was demonstrated by LH07 that the ratio of the maximal amplitudes of these torques, i.e. $q_{\mathrm{max}}=\Gamma_{1,\mathrm{max}}/\Gamma_{2,\mathrm{max}}$ is a key parameter which determines the dynamics of grain alignment. { This parameter depends on the grain composition and, similar to the grain cross-section on the ratio of wavelength to the effective grain size $\lambda/a$.}

	The LH07 study introduced the requirements for the existence of the high-$J$ attractors, $q_{\mathrm{max}}$ and the angle $\psi$ between the direction of radiation and magnetic field. For the parameter space where there are no high-$J$ attractor points, the RATs {\it decrease} the grain rotational velocities bringing grains in low-$J$ attractor points. 
	
\begin{figure}
		\centering
		\includegraphics[width=\linewidth]{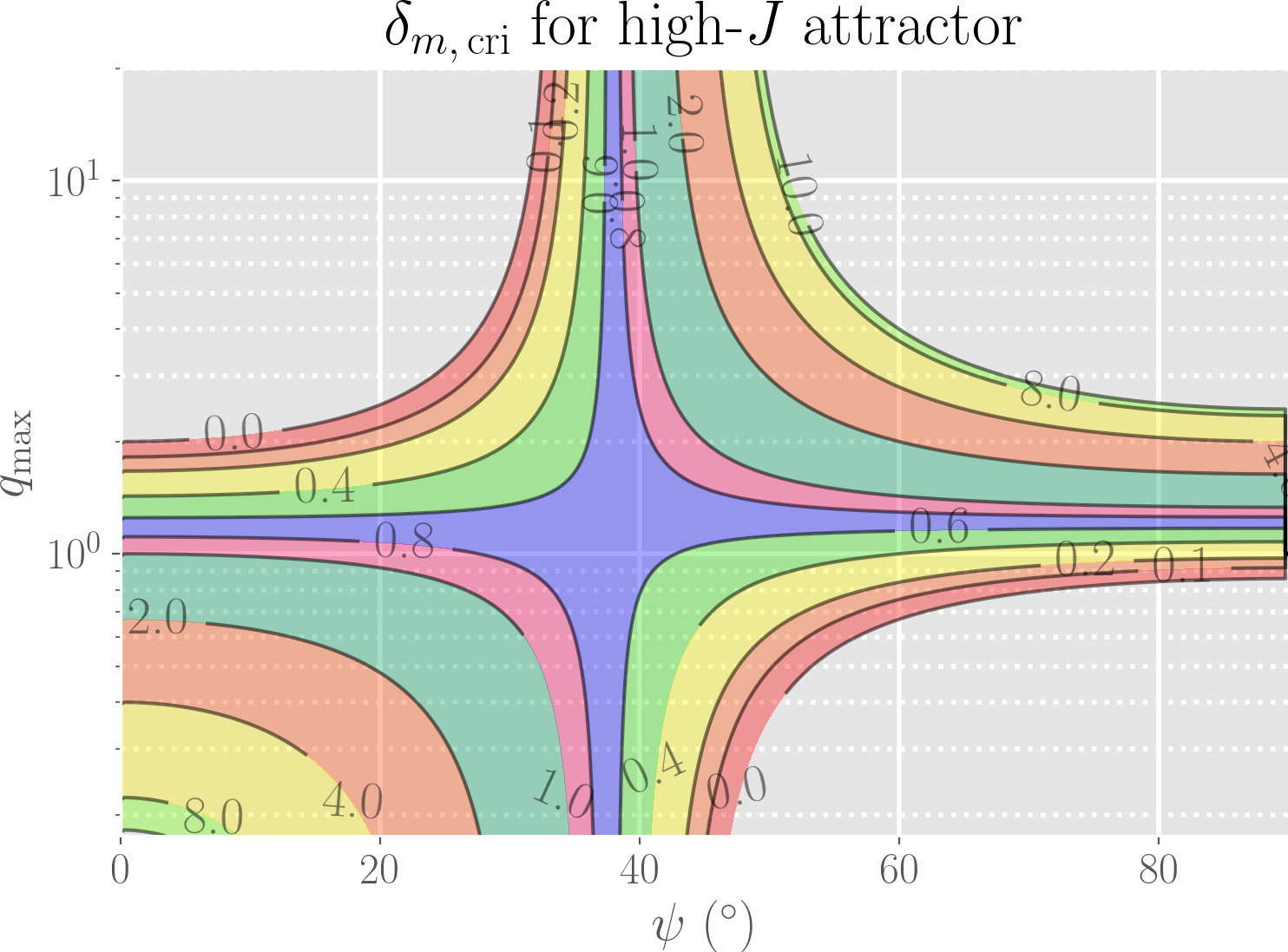}
		\caption{Contour map of critical magnetic inclusion ratio $\delta_{m}$ required for AMO to produce high-$J$ attractor points with respect to the q-factor $q_\mathrm{max}$ and angle $\psi$ between radiation and magnetic field directions. }
		\label{fig:figure1}
	\end{figure}

	While the RAT alignment does not require any effect of paramagnetic relaxation, it was demonstrated in \citet{Lazarian2007b} that an enhanced magnetic dissipation can stabilize high-$J$ stationary points, transferring them to high-$J$ attractor points. The effect was quantified in \citet{Hoang2016}, with the results reproduced in Figure \ref{fig:figure1} using idealized\footnote{Meaning the simplest form of equations reproducing torque components similar to ones considered in LH07 for a mirror at a 45$^\circ$ angle measured between the mirror normal and AMO grain intermediate axis, see LH07 for the exact definition.} AMO of a spinning grain with torque components (LH07)
	\begin{equation}\label{eq:AMO-components}
	\begin{aligned}
	\Gamma_1 &= \frac{q_\mathrm{max}}{3}(5\cos^2\Theta-2),\\
	\Gamma_2 &= \sin 2\Theta,\\
	\Gamma_3 &= 0.
	\end{aligned}	
	\end{equation} 
	Above, the angle $\Theta$ in scattering frame can be written in terms of the alignment frame angles $\xi$, $\psi$, and $\phi$ (see Figure \ref{fig:figure2}) as \citep{Draine1997}
	\begin{equation}\label{eq:thetaxipsiphi}
	\Theta = \arccos\left( \cos\xi\cos\psi - \sin\xi\sin\psi\cos\phi \right).
	\end{equation}
	
	Additionally, a factor  
	\begin{equation}
	\delta_m=\frac{\tau_{\mathrm{ran}}}{\tau_{m}},
	\label{delta}
	\end{equation}
	is introduced, where $\tau_{\mathrm{ran}}$ is the time of randomization by different processes that include gas bombardment, emission of photons, interaction with ions etc. \citep[see][for a complete list of randomizing processes]{Draine1998a} and $\tau_m$ is the time of the magnetic dissipation of the rotational energy of the grain. The parameter $\delta_m$ is a key factor for the Davis-Greenstein mechanism, but for the RAT alignment it plays an auxiliary role in increasing $f_{\mathrm{high}J}$.   	
	
	While the angle $\psi$ is determined by the relative position of the radiation source to the direction of grain precession about the magnetic field, $q_{\mathrm{max}}$ is an intrinsic property of grain determined by its shape and composition. Our work is intended to find out the percentage of realistic irregular grains that can have high-$J$ attractor points for different settings.
	
	\section{Sample of grains to study}\label{sec:grainmodel}
	Analysis of RATs on irregular grains is done using ensembles of Gaussian random ellipsoids \citep{Muinonen2011} (GRE). A single GRE shape can be identified by two statistical parameters, the correlation length between displacements and the standard deviation of displacements from the undeformed ellipsoid surface. Regeneration of a single explicit shape is possible by fixing the axial ratios of an ellipsoid and resetting the seed of random number generation. The method provides means of producing large numbers of random shapes, suitable for assessing the predictions of AMO, although GREs are unlikely to be a perfect representative of interstellar grain shapes.
	
	We consider ensembles of oblate and prolate grains of three different sizes, with equivolume spherical radii $a_\mathrm{eff}$ of 0.1 {\textmu}m ($N=1000$), 0.2 {\textmu}m ($N=1000$), and 0.75 {\textmu}m ($N=400$). The grains are composed of astronomical silicate \citep{Draine1984}, which, in the visual wavelength range has a complex refractive index close to $n=1.68+\mathrm{i}0.03$.
	
	The grain shapes are deformed from a smooth oblate/prolate spheroid to Gaussian random oblate/prolate spheroids (hereafter referred to as Gaussian oblate/prolate grains). The deformations for the basic shape, which have aspect ratios $a:b:c = 1:1:0.5$ for the oblate shape and $1:0.5:0.5$ for the prolate shape, all follow lognormal statistics for the radius deformation with a standard deviation 0.125, and a correlation length 0.35 between points on the spheroid.
	
	The Gaussian random deformation process provides means of studying statistical samples of random shapes with control over their general inertial properties using a minimal amount of parameters. The chosen aspect ratios reflect the assumption that high rotational speeds would likely disrupt \citep{Hoang2019, Hirashita2020} more extreme shapes.\footnote{The relation between the grain disruption and the grain alignment is explored in \citet{Lazarian2020b}. It is shown there that in some situations, e.g. in accretion disks, grains at high-J state of rotation can survive in the presence of intensive radiation. We, however, consider a more general case of interstellar disruption.}  and that sphere-like inertial and shape properties are the most unlikely to differ from results by Mie scattering analysis of spheroidal grains.
	
	For the sake of consistency, the statistical behavior of the deformed grain radii and moments of inertia are studied. The principal moments of inertia for an oblate and a prolate spheroid are
	\begin{equation}
	\begin{cases}
	I_{p,\mathrm{oblate}} = \dfrac{m}{20}\:\mathrm{diag}(5,5,8),\\
	I_{p,\mathrm{prolate}} = \dfrac{m}{20}\:\mathrm{diag}(2,5,5),
	\end{cases}
	\end{equation}
	where diag denotes elements on a matrix diagonal. We compare the statistical average and standard deviation of diagonalized inertia matrix components to determine whether the inertial properties of the base shape are conserved. We also compare the deformed volumetric radii $R_{\mathrm{vol}}$ (also radius of the sphere of equivalent volume, or $a_\mathrm{eff}$) with those of the base shapes. A distinction between the two is made here, as in the scattering calculations all the grains are scaled so that a desired $ a_{\mathrm{eff}} $ is achieved, whereas here we use $ R_{\mathrm{vol}} $ to measure with a single number whether or not the shape deformation affects the overall shape. For a general ellipsoid, the volumetric radius is simply $ R_{\mathrm{vol}} = (abc)^{1/3}$. The results are collected in Table \ref{table:deformstatistics}. We see that the randomization process indeed produces expected values close to that of the base shapes.
	
	\begin{table*}
		\caption{Ensemble properties compared to the undeformed base shapes. The dimensionless units of length are chosen so that the semimajor axes have values 1 and semiminor axes have values 0.5. The components of $ I_{p} $ and the corresponding standard deviations have been scaled so that the last components, or maximal moments of inertia, are equal. Correlation length is fixed at 0.35 for all shapes.}
		\begin{tabular}{r||c|c|c||c|c|c}\label{table:deformstatistics}
			& $R_{\mathrm{vol}}$ & E{[}$R_{\mathrm{vol}}${]} & $\sigma$ & diag$(I_{p})$ & E{[}diag$(I_{p}){]}$ & $\sigma$ \\
			\hline
			Oblate & 0.7937 & 0.8026
			& 0.0482 & $ (5,5,8) $ & $ (4.185, 5.899, 8) $ & $ (0.355, 1.183, 1.101) $ \\
			Prolate & 0.6300 & 0.6454 & 0.0582
			& $ (2,5,5) $ & $ (1.978, 4.578, 5) $ & $ (0.306, 1.009, 0.954) $ 
		\end{tabular}
	\end{table*}
	
	\section{Alignment fraction analysis for Gaussian random ensembles}\label{sec:calculations}
	The GRE ensemble as a starting point, normalized radiative torques are solved using the $T$-matrix method. As the $T$-matrix describes grain scattering properties at a single wavelength, the following results are calculated for a 10-element sample of a wavelength range $\lambda \in [300,1920]$ nm, equally divided between the endpoints.
	
	For maximal applicability of the results in astronomical context, the interstellar radiation field (ISRF) of local solar neighborhood \citep{Mathis1983} is hereafter assumed as the distribution of incident energy density, unless explicitly wavelength dependent quantities are considered. When RATs under ISRF illumination are considered, they are computed using the definition
	\begin{equation}\label{eq:ISRF-RAT-definition}
	Q = \frac{\int Q_\lambda u_\lambda\mathrm{d}\lambda}{\int u_\lambda \mathrm{d}\lambda} = \frac{\int Q_\lambda u_\lambda\mathrm{d}\lambda}{u_\mathrm{rad}}.
	\end{equation}
	
	\subsection{Equations of motion for alignment}\label{sec:f-analysis}
\begin{figure}
		\centering
		\includegraphics[width=\linewidth]{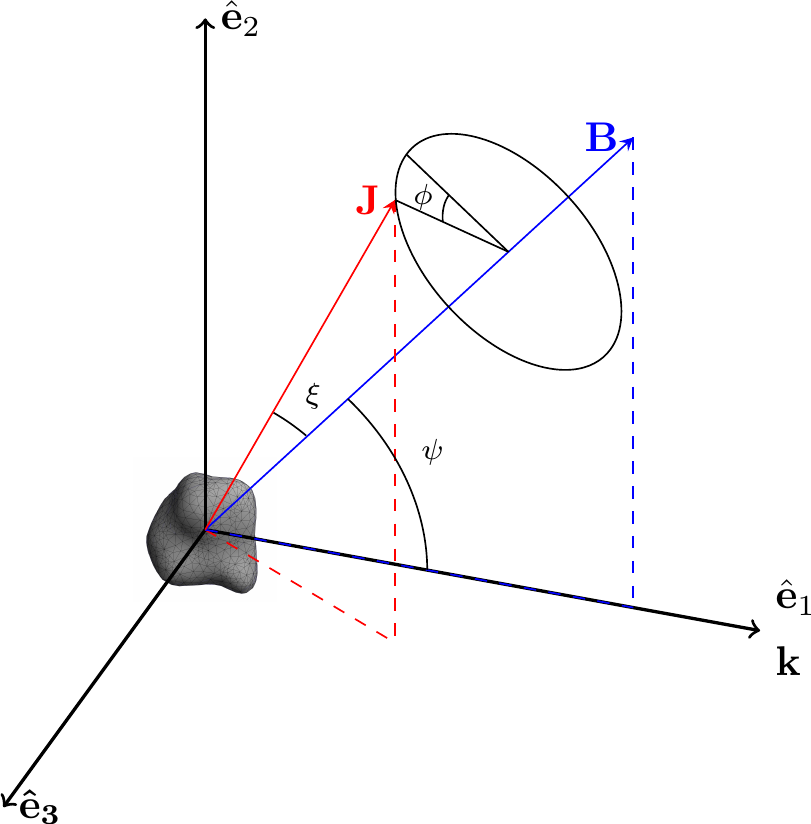}
		\caption{Alignment coordinate system.}
		\label{fig:figure2}
	\end{figure}

	We work in the alignment coordinates $(\xi,\psi,\phi)$, where $\xi$ is the alignment angle between angular momentum and magnetic field vectors $\mathbf{J}$ and $\mathbf{B}$, $\psi$ the angle between $\mathbf{B}$ and anisotropic radiation direction $\mathbf{k}$, and $\phi$ the Larmor precession angle of $\mathbf{J}$ around $\mathbf{B}$. The coordinate system is illustrated in Figure \ref{fig:figure2}. In the following analysis, averaging a quantity $q$ over $\phi$ is denoted by $\bar{q}$. 
	
	Using the equations of motion of RATs where perfect internal alignment is assumed, we have, similarly as in LH07, for the alignment angle $\xi$ and angular momentum $J'= J/I_{1}\omega_{T}$ \citep{Hoang2016}: 
	\begin{equation}\label{eq:eoms1}
	\frac{\mathrm{d}\xi}{\mathrm{d}t'} = \frac{M}{J'}\bar{F}(\xi,\psi) - \delta_{m}\sin\xi\cos\xi,
	\end{equation}
	\begin{equation}\label{eq:eoms2}
	\frac{\mathrm{d}J'}{\mathrm{d}t'} = M\bar{H}(\xi,\psi) - J'(1+\delta_{m}\sin^{2}\xi).
	\end{equation}
	Above, $ t' = t/\tau_{\mathrm{gas}}$, $ I_{1} $ is the maximal grain moment of inertia, $ \omega_{T} = (2k_{B}T_{\mathrm{gas}}/I_{1})^{1/2} $, $ M = \gamma\bar{\lambda}u_{\mathrm{rad}}a_{\mathrm{eff}}^{2}/2$. Finally, $ \bar{F} $ and $ \bar{H} $ are the $ \phi $-averaged aligning and spin-up RAT components.
	
	The stationary points $ (\xi_{s}, J_{s}) $ can be found setting the equations of motion to zero, which leaves us with
	\begin{equation}\label{eq:statpoints1}
	\frac{\bar{F}(\xi_{s})(1+\delta_{m}\sin^{2}\xi_{s})}{\bar{H}(\xi_{s})} = \delta_{m}\sin\xi_{s}\cos\xi_{s}.
	\end{equation}
	The zeros of equation (\ref{eq:statpoints1}) now give, as shown in LH07, the stationary points in the $(\xi, J)$ phase space. Specifically, we find universal stationary points $ \sin\xi_{s} = 0$, which have, after $\phi$-averaging, $ \bar{F} = 0$. 
	
	Generally, for a stationary point to be an attractor, we have the requirements, as given by \citet[DW97]{Draine1997},
	\begin{equation}\label{statpoints2}
	A+D<0,\,BC-AD<0,
	\end{equation}
	where 
	\begin{equation}
	\begin{aligned}
	A &= \frac{M\nabla\bar{F}}{J'_{s}}-\delta_{m}\cos{2\xi_s} ,\, B = -\frac{M}{J'^2_{s}}\bar{F}, \\
	C &= M\nabla\bar{H}-\delta_{m}J'_{s}\sin{2\xi_{s}},\, D = -(1+\delta_{m}\sin^2\xi_{s}), 
	\end{aligned}
	\end{equation}
	and $ \nabla \equiv \frac{\mathrm{d}}{\mathrm{d}\xi} $. When we consider the universal stationary points, the requirement for high-$J$ attractors simplifies to
	\begin{equation}\label{simpleattractor}
	\frac{\nabla\bar{F}}{\bar{H}}\bigg|_{\sin\xi_{s}=0} -\delta_{m} < 0, \quad \bar{H}>0. 
	\end{equation}
	
	When the equations of motion are analysed for grain ensembles as a function of $\psi$, we can find $f_{\mathrm{high}J}$. In the following calculations, we assume that the maximal axes of inertia of grains are aligned with the magnetic field, i.e. the universal stationary points only are considered. Physically, this corresponds to a situation where the alignment process has reached a steady state, and all alignable grains are aligned. Thus, the fraction $f_{\mathrm{high}J}$ is the fraction of grains for which the requirement (\ref{simpleattractor}) for any of the two universal stationary points is satisfied.
	
	\subsection{Distributions of $q_\mathrm{max}$ of GRE grains}
\begin{figure*}
		\centering
		\includegraphics[width=\textwidth]{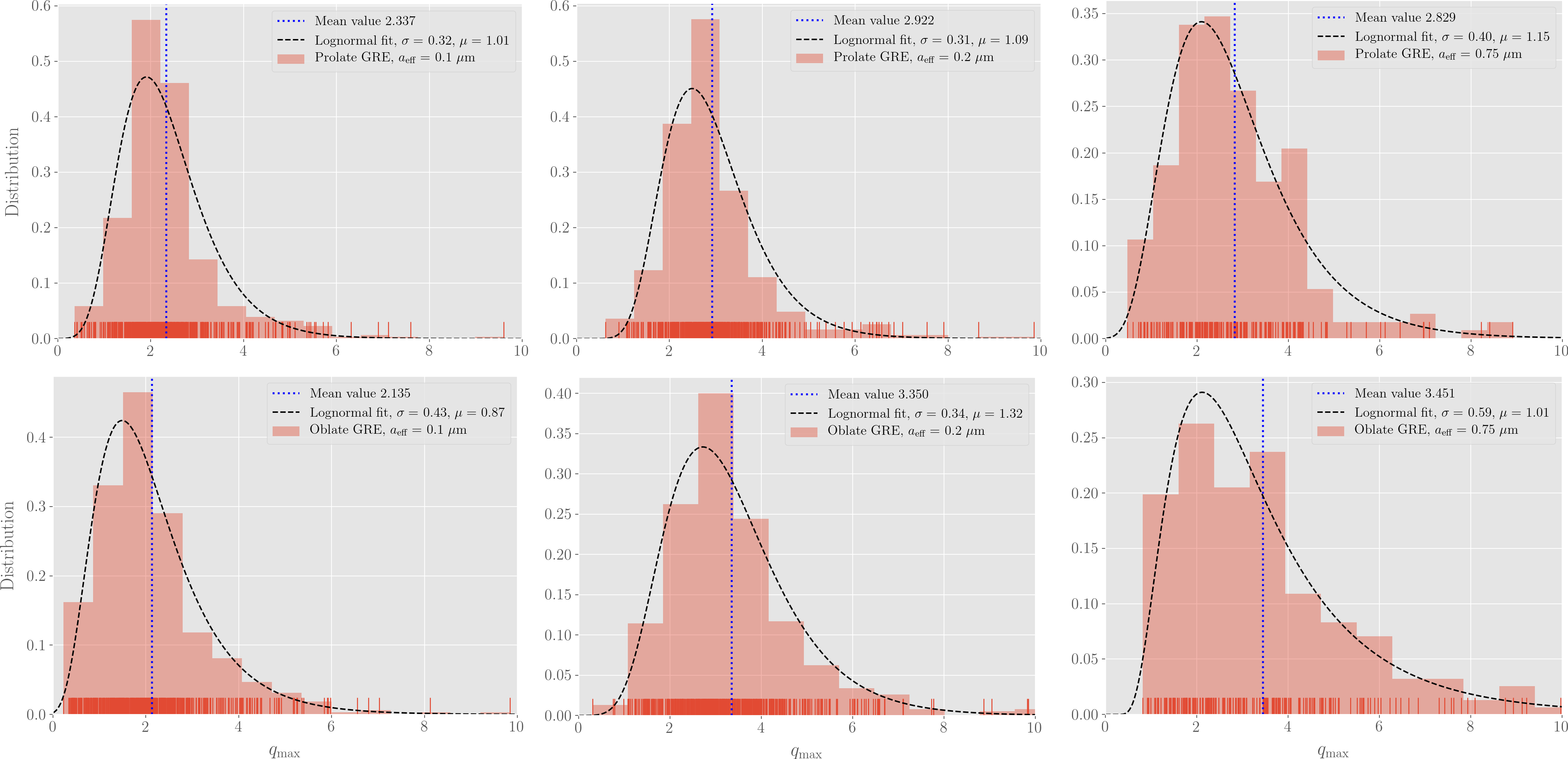}
		\caption{Distributions of $q_\mathrm{max}$ values along with lognormal fits for each subensembles of separate size and base shape.}
		\label{fig:figure3}
	\end{figure*}

	In order to derive predictions for alignment degrees with AMO, a distribution of $q_\mathrm{max}$ corresponding to some dust grain population is required. Now, we focus on $q_\mathrm{max}$ distributions obtained from the GRE ensemble. In Figure \ref{fig:figure3}, distributions of $q_\mathrm{max}$ for the three differently sized oblate and prolate GREs are presented, along with lognormal fits for the distributions. 
	
	Lognormal distribution was chosen to provide the fits for the adequate quality of the fit and the fact that the generation of GRE shapes involve lognormal statistics. It should be noted that no claims on the connection between the shape of the $q_\mathrm{max}$ distribution and the underlying GRE statistics are made, and such analysis is well out of the current scope.
	
	The lognormal distribution, which can be generated using logmean $\mu$ and logdeviation $\sigma$ as parameters, provides adequate fits for all the separate-size-shape grain populations. Further, for the total ensemble of 2400 grains the lognormal distribution provides a good fit with $\sigma=0.37$, $\mu=1.17$ (see Figure \ref{fig:figure10}, right panel). 
	
\begin{figure}
		\centering
		\includegraphics[width=\linewidth]{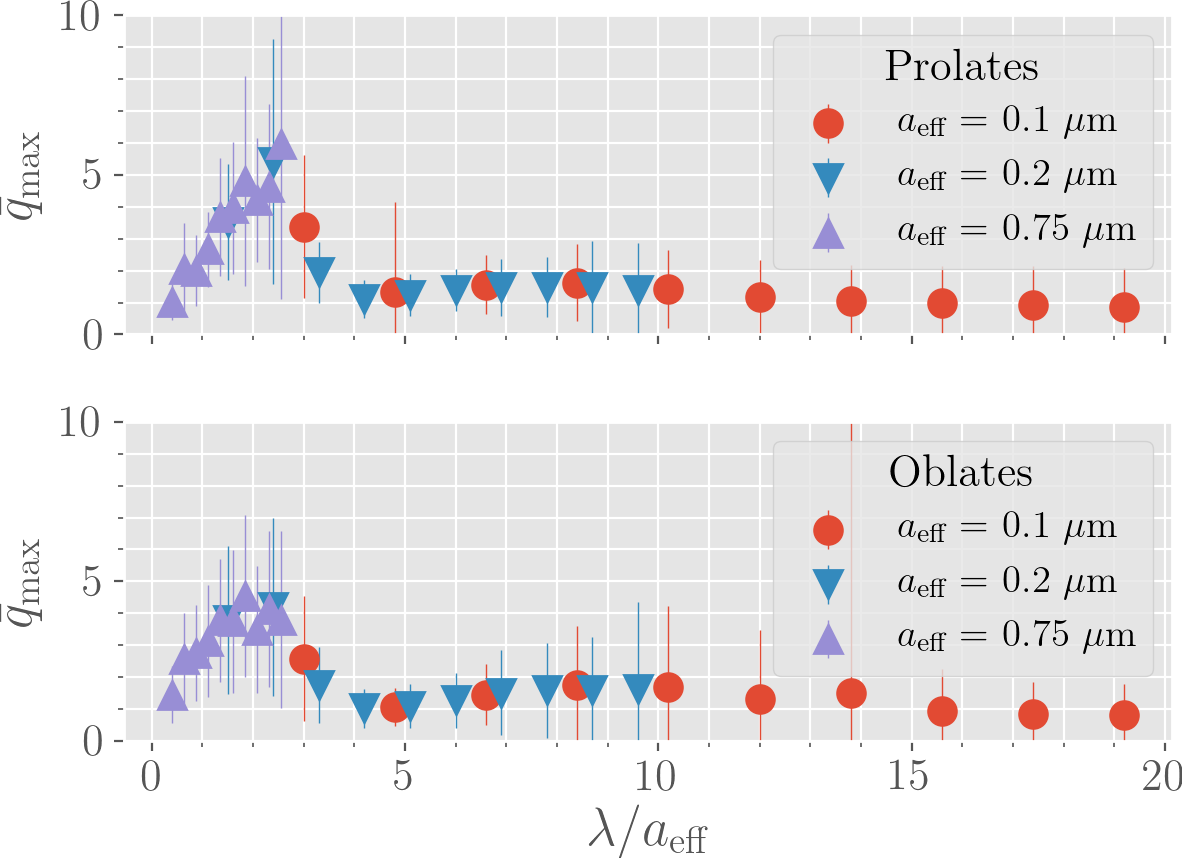}
		\caption{Average $\bar{q}_\mathrm{max}$ as a function of $\lambda/a_\mathrm{eff}$. Vertical bars represent one standard deviation.}
		\label{fig:figure4}
	\end{figure}

	The mean value of the subensemble $q_\mathrm{max}$ distribution increases with the grain size in the ISRF. For extrapolations, it is important to identify the functional form of the average $q_\mathrm{max}$ with respect to the ratio $\lambda/a_\mathrm{eff}$ of wavelength and grain size. In Figure \ref{fig:figure4}, the mean values are presented for the oblate and prolate subensembles.
	
	We identify a significant peak in $\bar{q}_\mathrm{max}$ (bar indicating an ensemble mean) at $\lambda/a_\mathrm{eff}=2$ for both the prolate and oblate subensembles. In terms of grain sizes, there thus will be, for any given incident energy density distribution, some grain size for which a maximal $\bar{q}_\mathrm{max}$ is obtained. Now, the ISRF according to \citet{Mathis1983} peaks at 700 nm, and the UV peak at 120 nm is left outside the wavelength range considered.
	
	Extrapolating similar functional form of $\bar{q}_\mathrm{max}$ for any grain size, the peak location will correspond to grain size of 0.35 {\textmu}m. Therefore, for the ISRF, the distribution of $q_\mathrm{max}$ for grains with $a_\mathrm{eff} = $ 0.35 {\textmu}m would likely tend most towards large values.
	
	Additionally, we note that $\bar{q}_\mathrm{max}\approx 1$ is a local minimum between $\lambda/a_\mathrm{eff}=4$ and $5$ with a low standard deviation, especially in the oblate case. Therefore, at this size range, we expect the ensemble behaviour to be one of the most representative of the difference between irregular grains and AMO, which predicts no high-$J$ attractors without magnetic inclusions at $q_\mathrm{max}=1$.
	
	The distributions of $q_\mathrm{max}$ have mean values close to $q_\mathrm{max}=3$. Considering AMO, using Figure \ref{fig:figure1}, we predict that if most grains have $q_\mathrm{max}$ values close to 2 or above, the fraction $f_{\mathrm{high}J}$ will fall quickly to zero at around $\psi=30^\circ$. Next, we test how the GRE ensemble results compare to the AMO prediction.
	
	\subsection{Ensemble high-$J$ fractions $f_{\mathrm{high}J}$}
	We compute the high-$J$ alignment fractions for the GRE subensembles using the procedure described in Section \ref{sec:f-analysis}. For simplicity, magnetic inclusions are not considered, i.e., $\delta_{m}=0$.
	
\begin{figure*}
		\centering
		\includegraphics[width=0.9\textwidth]{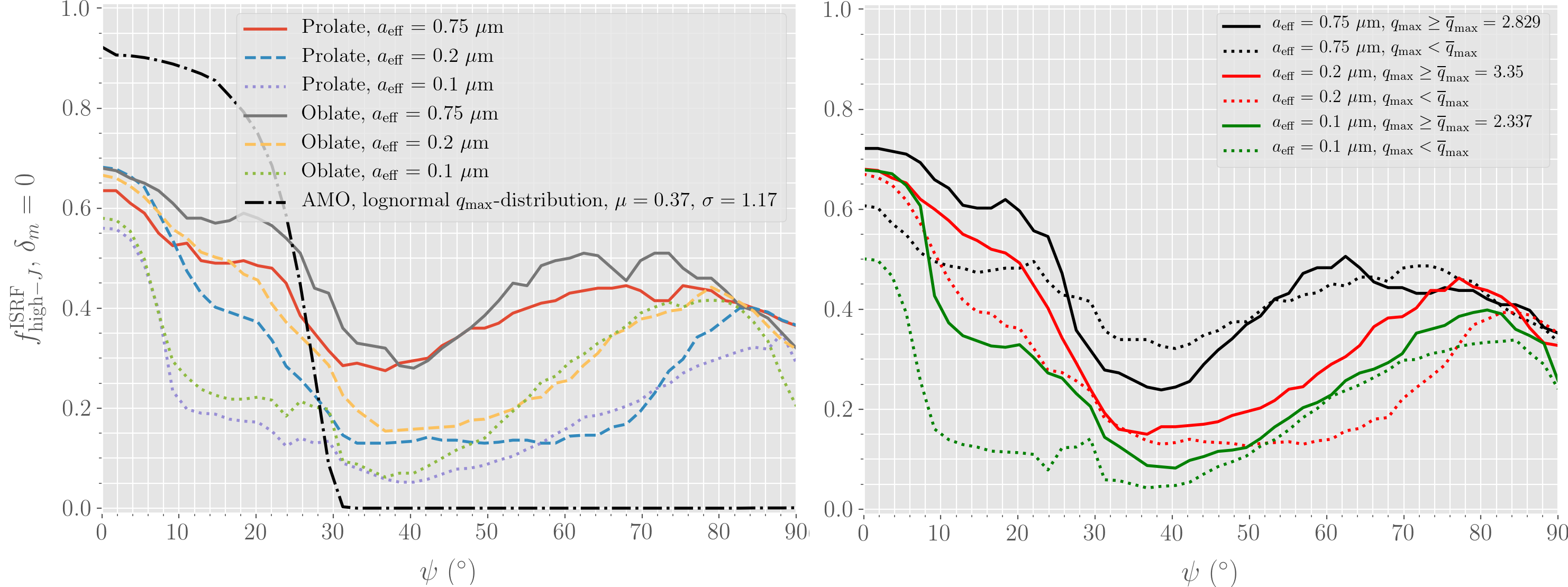}
		\caption{Left panel: Fraction $f_{\mathrm{high}J}$ of grains with high angular momentum attractors of the six different GRE subensembles and AMO, as a function of the angle $\psi$ between magnetic field and radiation field directions. Right panel: Similar to the left panel, but for a reorganization of the ensembles according to grain size and the $q_\mathrm{max}$ value compared to the average $q_\mathrm{max}$ of each different grain sizes.}
		\label{fig:figure5}
	\end{figure*}

	First, in the left panel of Figure \ref{fig:figure5}, the $f_{\mathrm{high}J}$ of the subensembles are compared against the prediction of AMO with a lognormal ($\sigma=0.37$, $\mu=1.17$) $q_\mathrm{max}$ distribution. As $\delta_{m}=0$, the $f_{\mathrm{high}J}$ of AMO falls to zero just after $\psi$ has reached $30^\circ$, as is predictable from Figure \ref{fig:figure1} for a distribution of $q_\mathrm{max}$ that does not contain grains with $q_\mathrm{max}<0.5$.
	
	GRE ensembles exhibit high-$J$ attractors more universally than predicted by AMO. Two important observations can immediately be identified from the data: First, $f_{\mathrm{high}J}$ both increases and becomes more uniform with respect to $\psi$ along with increasing grain size. Second, for all subensembles, the minimum value of $f_{\mathrm{high}J}$ is centered around $\psi = 40^\circ$, near the angular range where AMO predicts no high-$J$ attractors without paramagnetic inclusions, and smaller the grain, wider the minimum. Most grain types also exhibit a local maximum around $\psi = 70^\circ$ -- $80^\circ$.
	
	The difference between $f_{\mathrm{high}J}$ of the oblate and prolate grains is slightly in the favour of oblate grains. The difference between base shape types is mostly less than 10 \%, with the exception of the 0.2 {\textmu}m cases in the approximate range $\psi\in(60^\circ,70^\circ)$.
	
	Second, in the right panel of Figure \ref{fig:figure5}, a comparison of rearranged subensembles is performed. Rearrangement is done so that grains of certain size with $q_\mathrm{max}$ values less or greater than the mean $q_\mathrm{max}$ value of all grains at the said size are separated to different subensembles.
	
	According to Figure \ref{fig:figure1}, grains with large $q_\mathrm{max}$ are expected to have higher $f_{\mathrm{high}J}$ at small $\psi$, and lower $f_{\mathrm{high}J}$ large $\psi$, when compared to the counterpart with small $q_\mathrm{max}$ values. However, we see that this is not the case for irregular grains. Instead, grains with larger $q_\mathrm{max}$ have almost universally at least as large $f_{\mathrm{high}J}$ as the counterpart with smaller $q_\mathrm{max}$, the most notable exception being the 0.75 {\textmu}m case around $\psi=40^\circ$.
	
\begin{figure*}
		\centering
		\includegraphics[width=0.9\textwidth]{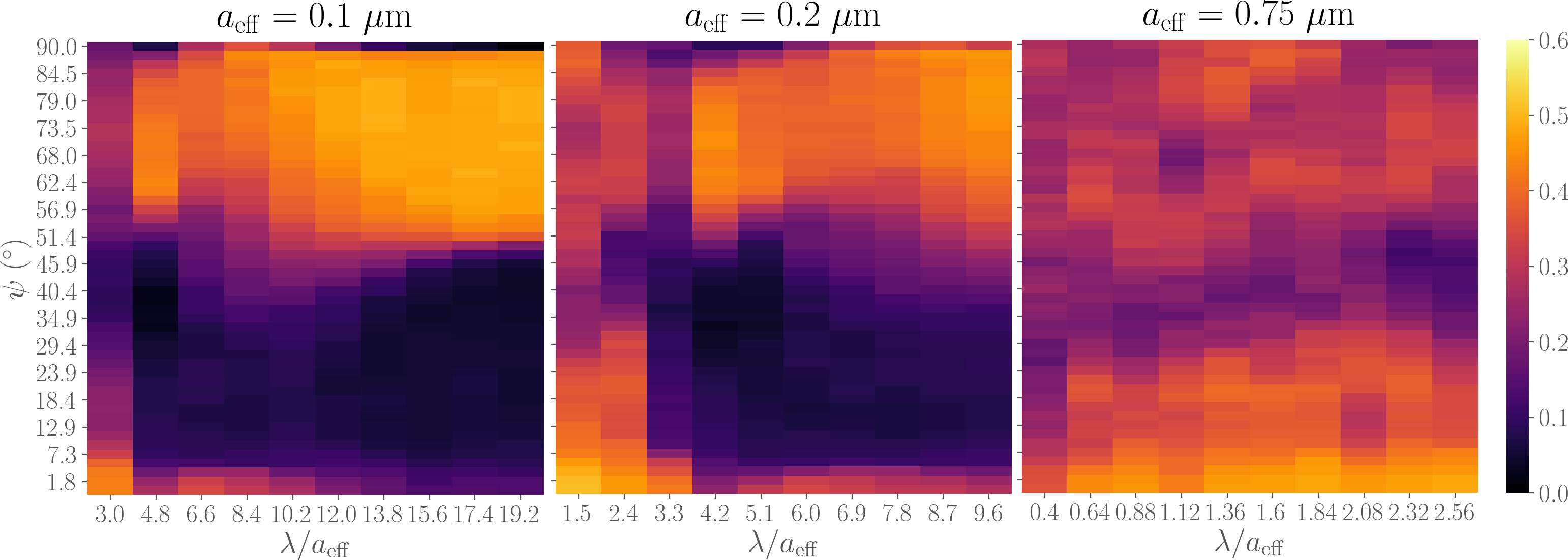}
		\caption{Fraction $f_{\mathrm{high}J}$ as a function of $\lambda/a_\mathrm{eff}$ and $\psi$ for the 3 oblate subensembles. All subfigures share the same color scaling.}
		\label{fig:figure6}
	\end{figure*}

\begin{figure*}
		\centering
		\includegraphics[width=0.9\textwidth]{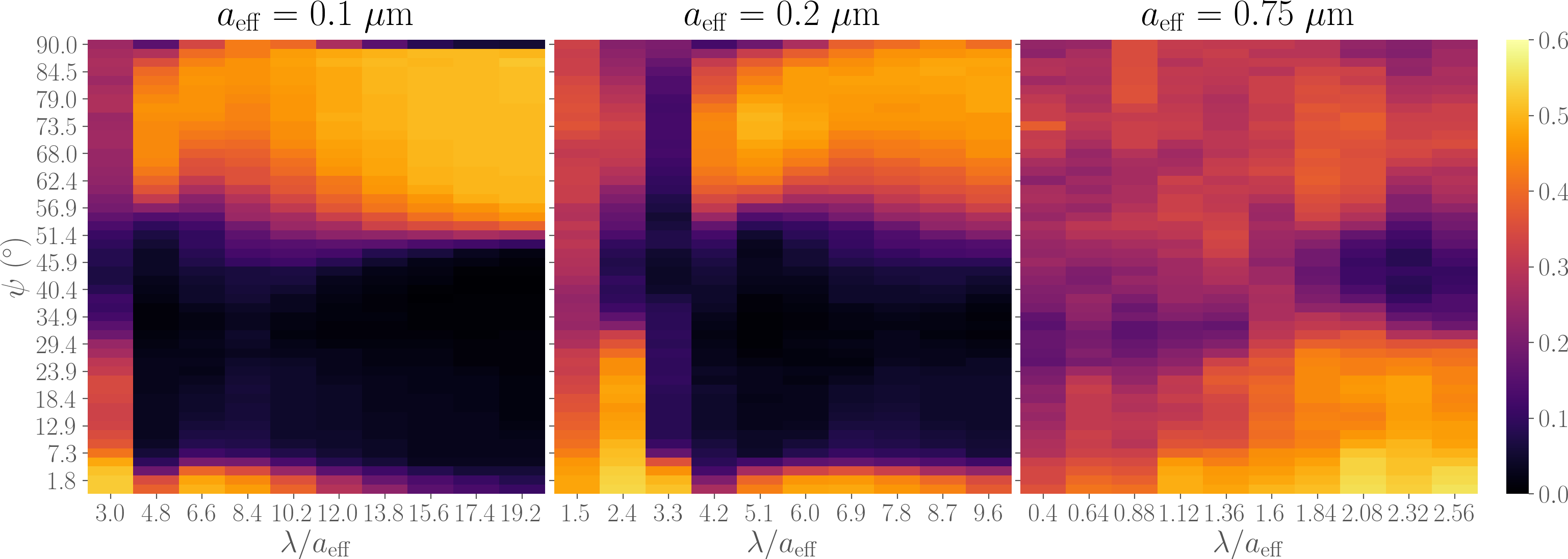}
		\caption{Same as Figure \ref{fig:figure6} for the 3 prolate subensembles.}
		\label{fig:figure7}
	\end{figure*}

	Third, in order to identify the contribution of RATs by different wavelengths on $f_{\mathrm{high}J}$, Figures \ref{fig:figure6} and \ref{fig:figure7} show $f_{\mathrm{high}J}$ as a function of $\lambda/a_\mathrm{eff}$ and $\psi$.	
	
	Comparing the columns of Figures \ref{fig:figure6} and \ref{fig:figure7} to the functional form of $\bar{q}_\mathrm{max}$ confirms that a high value of $q_\mathrm{max}$ is required to produce high-$J$ attractors at small $\psi$. Furthermore, as $\bar{q}_\mathrm{max}$ of both the oblate and prolate cases are highly similar, so are Figures \ref{fig:figure6} and \ref{fig:figure7}, also.
	
	Next, Figures \ref{fig:figure6} and \ref{fig:figure7} can be compared against Figure \ref{fig:figure1} with the help of $\bar{q}_\mathrm{max}$ data from Figure \ref{fig:figure4}. At $\lambda/a_\mathrm{eff}\approx 5$, where $\bar{q}_\mathrm{max}$ is at a minimum, the contour structure of Figure \ref{fig:figure1} predicts that the cutoff angle $\psi$ for which high-$J$ begins to occur, should also be at a minimum. Such slight drift is visible in both Figures \ref{fig:figure6} and \ref{fig:figure7}. Additionally, at the middle of the cross-shaped contour structure, no high-J attractors should exist without magnetic inclusions. For AMO, the high-$J$ poor region occurs at $q_\mathrm{max}\approx 1.2$, while for irregular grains, the region of least high-$J$ attractors occurs at $\lambda/a_\mathrm{eff}\approx 3$, or at $q_\mathrm{max}\approx 2$, or 3 (for small prolates).
	
	As a last observation, the 0.75 {\textmu}m cases showcase well the discrepancy between $f_{\mathrm{high}J}$ of AMO and the GRE ensemble for $\psi<40^\circ$. As shown in Figure \ref{fig:figure4}, only a small fraction of the grains have $q_\mathrm{max}<1$, yet there exists more high-$J$ attractor points at values of $\psi$ larger than $40^\circ$.
	
	It is important to note that because $q_\mathrm{max}$ is not a free parameter for irregular grains, and because the irregular grain radiative torques can have highly variable functional form for two grains with similar $q_\mathrm{max}$, a figure such as Figure \ref{fig:figure1} is difficult, if not impossible, to be produced for the GRE ensemble. Thus, comparison of the radiative torques of AMO and of the GRE ensemble is crucial in order to explain the discrepancy between $f_{\mathrm{high}J}$ of AMO and the GRE ensemble of Figure \ref{fig:figure5}.
	
	\subsection{Differences between GRE ensembles and AMO}
	Potentially the main explanation for the differences between obtained GRE ensemble results and the predictions derived from AMO is the different functional forms of the radiative torques themselves. In this section, we compare the radiative torque components in both the scattering frame and alignment frame. In both cases, the mean square differences $\left\langle \Delta^2 \right\rangle q$ of GRE and AMO quantities $q$,
	\begin{equation}\label{eq:MSE}
	\left\langle \Delta^2 \right\rangle q = \frac{1}{N} \sum_{i=1}^{N}(q_\mathrm{GRE}-q_\mathrm{AMO})^2,
	\end{equation}
	are determined.	
	
\begin{figure}
		\centering
		\includegraphics[width=\linewidth]{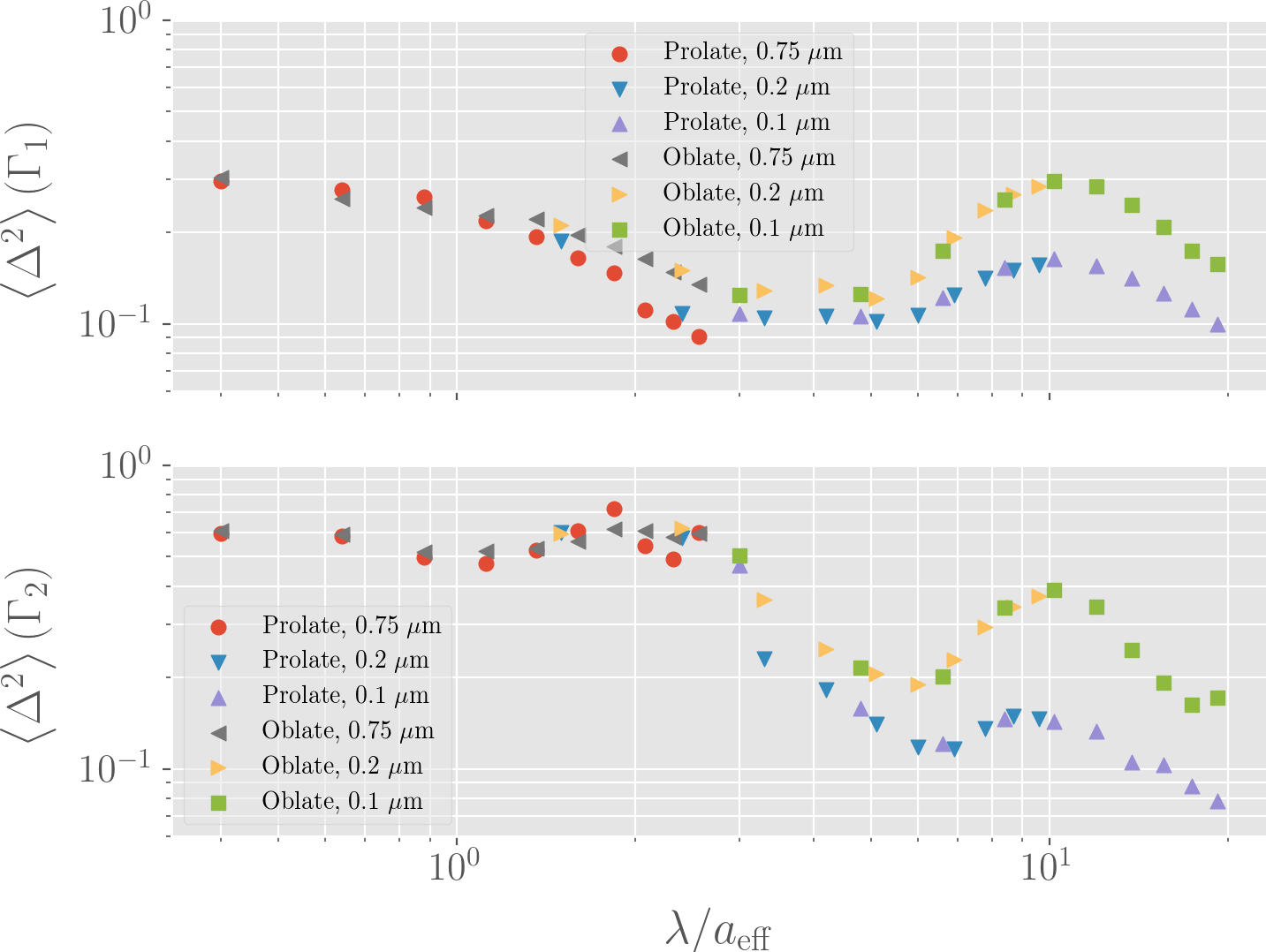}
		\caption{Mean square differences over angle $\Theta$ of scattering frame RAT components $\Gamma_1$ and $\Gamma_2$ between AMO and the GRE subensembles as a function of $\lambda/a_\mathrm{eff}$.}
		\label{fig:figure8}
	\end{figure}

	First, in Figure \ref{fig:figure8}, the shapes of scattering frame RAT components $\Gamma_1$ and $\Gamma_2$ are compared. In the procedure, the AMO RAT is produced according to the $q_\mathrm{max}$ of each grain in the GRE ensemble and normalizing the torques with the maximum values of $\Gamma_1$. The mean square difference is then calculated over the angle $\Theta$, and the result is presented as a function of $\lambda/a_\mathrm{eff}$. 
	
	According to the Figure \ref{fig:figure8}, AMO is generally more likely to produce similar RATs to GRE grains the smaller the grains are. The result is consistent with the results of Figures \ref{fig:figure6} and \ref{fig:figure7}, which show that smaller grains exhibit $f_{\mathrm{high}J}$ values closer to those predicted by AMO.
	
	Additionally, the difference $\left\langle \Delta^2 \right\rangle \Gamma_2$ reaches a peak value around $\lambda/a_\mathrm{eff} = 2$. The difference for $\Gamma_2$ remains large for larger grains, and for $\Gamma_1$, increases scrictly as grains grow larger. Therefore, for large grains, predictions of AMO can be deemed relatively inaccurate.
	
\begin{figure}
		\centering
		\includegraphics[width=\linewidth]{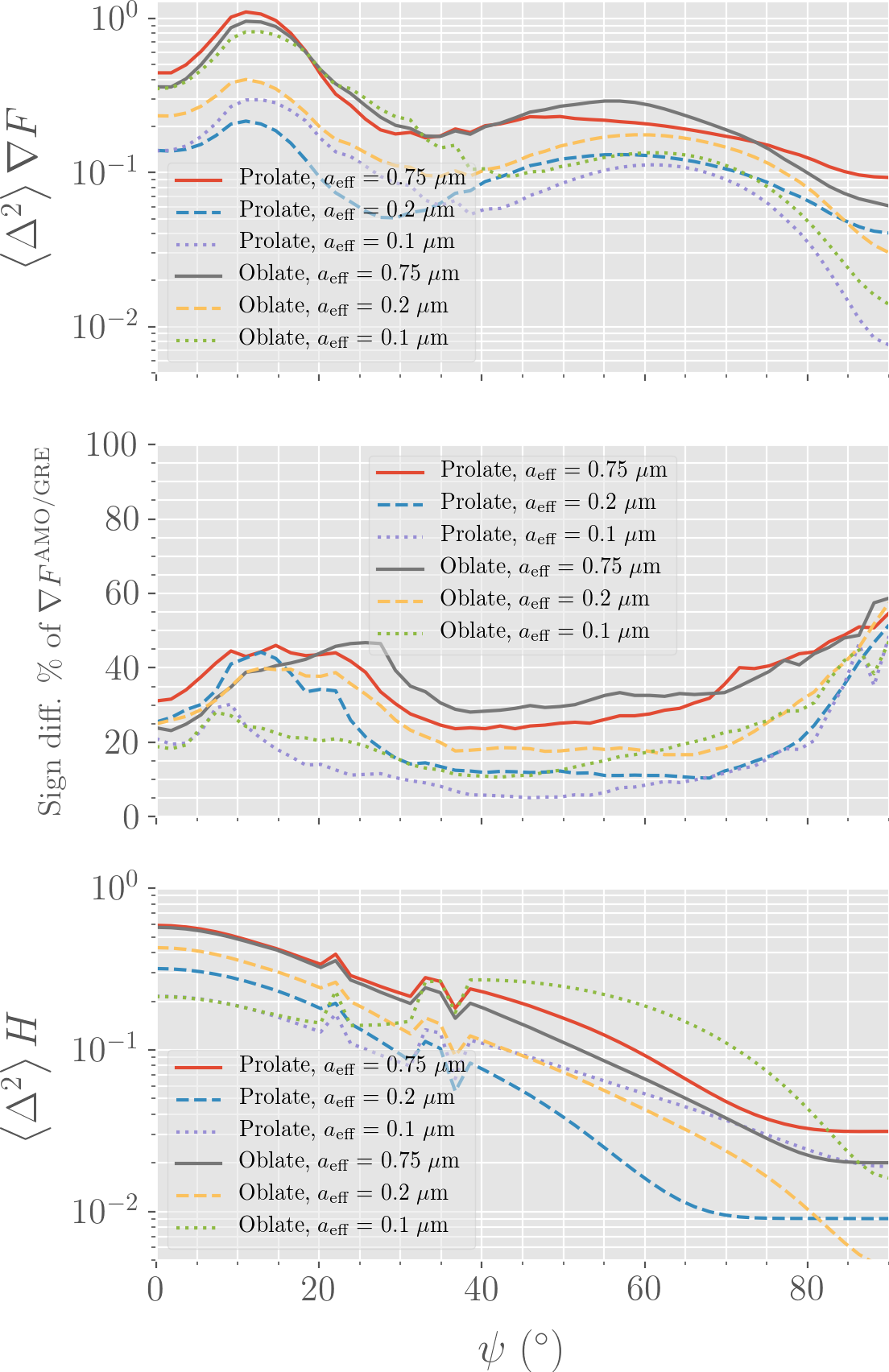}
		\caption{Mean square differences in the alignment frame over $\lambda/a_\mathrm{eff}$ of $\nabla F$ (top panel) and $H$ (bottom panel) between AMO and the GRE subensemble as a function of the external alignment angle $\psi$ between the magnetic field and radiation directions. Middle panel: The percentage of GRE subensembles with different sign of $\nabla F$ at $\sin\xi=0$ as a function of $\psi$}
		\label{fig:figure9}
	\end{figure}

	Next, alignment frame RATs are compared in order to potentially identify critical differences between AMO prediction of Figure \ref{fig:figure1} and $f_{\mathrm{high}J}$ of the GRE subensemble. The differences are determined over the whole GRE ensemble as a function of the angle $\psi$. Additionally, the sign difference percentages of $\nabla F$ between AMO and the GRE subensembles are determined. The results are collected in Figure \ref{fig:figure9}.
	
	Irregular grain RATs, particularly $\nabla F$ and its sign, differ from those of AMO at large values of $\psi$. The spin-up component $H$ appears to approach the one of AMO as $\psi$ approaches $90^\circ$. 
	
	Differences of $\nabla F$ are locally minimal at values around $\psi=40^\circ$, where $f_{\mathrm{high}J}$ reaches the minimal value with both AMO and the GRE subensembles (Figure \ref{fig:figure5}, left panel). 
	
	Sign difference of AMO and the GRE subensembles reaches up to 50\% at $\psi=90^\circ$. However, $f_{\mathrm{high}J}$ tends to decrease at $\psi>80^\circ$ for all grains to as low as 0.2, implying that the attractors at this region are low high-$J$ ones.
	
	To summarize the section, RAT differences show expected size-dependence that is evident from the $f_{\mathrm{high}J}$ results. Particularly differences in the scattering frame component $\Gamma_2$, which is expected to be important near $\psi=90^\circ$ (where radiation direction is perpendicular to $\mathbf{J}$) and the differences of the $\xi$-derivative of aligning RAT component $F$ are likely to be related to large $f_{\mathrm{high}J}$ values not predicted by AMO at large $\psi$.
	
	\subsection{Effect of magnetic inclusions}
\begin{figure*}
		\centering
		\includegraphics[width=\textwidth]{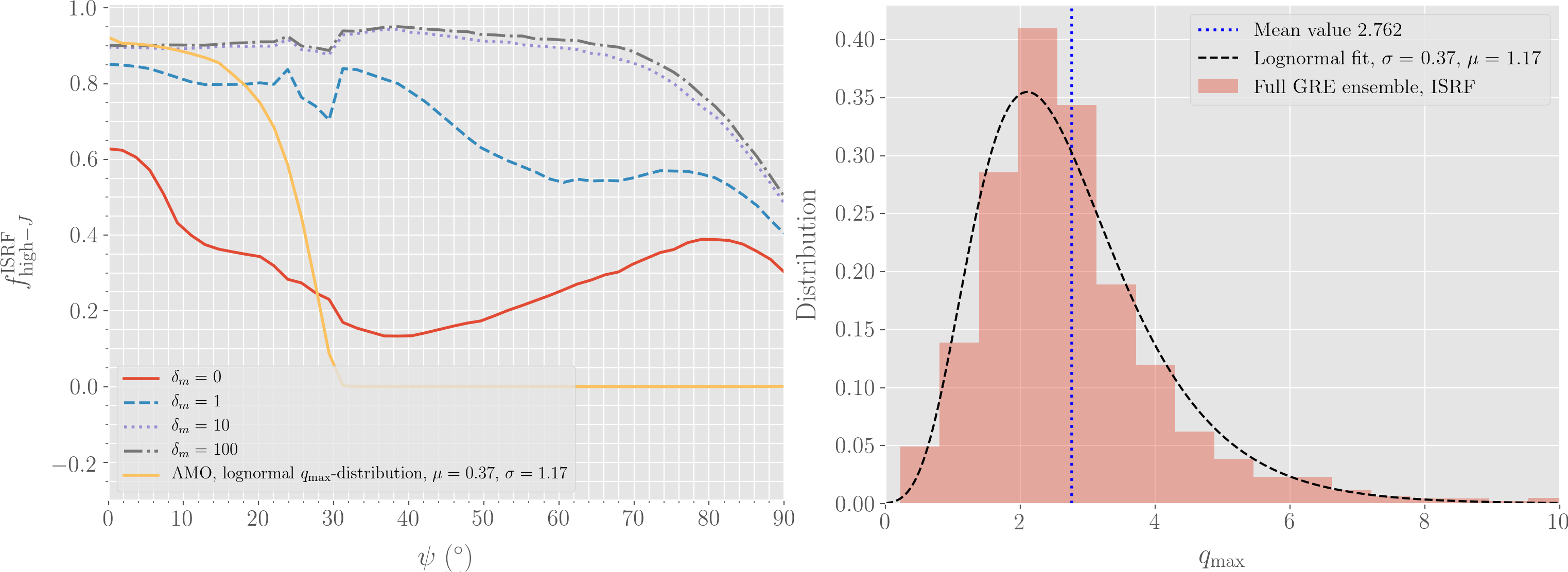}
		\caption{Left panel: Values of $f_{\mathrm{high}J}$ as a function of $\psi$ for different values of $\delta_m$ describing different relative importances of magnetic inclusions. AMO with $\delta_m=0$ for reference. Right panel: Distribution of $q_\mathrm{max}$ for the total GRE ensemble.}
		\label{fig:figure10}
	\end{figure*}

	According to Figure \ref{fig:figure1}, high-$J$ attractors are universal with moderate magnetic inclusions expect in the limit of large $q_\mathrm{max}$ and $\psi$. We test the existence of high-$J$ attractors when magnetic inclusions are added for the total GRE ensemble in Figure \ref{fig:figure10}.
	
	Increasing $\delta_m$ to 10 increases $f_{\mathrm{high}J}$ to 0.9 when $\psi<70^\circ$, i.e. high-$J$ attractors become universal at $\psi<70^\circ$. Above $\psi=70^\circ$, $f_{\mathrm{high}J}$ falls down to 0.5, as seen in the left panel of Figure \ref{fig:figure10}.
	
	Figure \ref{fig:figure1} predicts that at $\psi=90^\circ$, for grains with $q_\mathrm{max}>2.5$, high-$J$ attractors do not exist without extreme magnetic inclusions, if at all. According to the right panel of Figure \ref{fig:figure10}, the mean value of $q_\mathrm{max} = 2.762$. Thus, approximately half of the grains have $q_\mathrm{max}>2.5$, which makes the result in the left panel of Figure \ref{fig:figure10} consistent with Figure \ref{fig:figure1} and the conclusion that in the limit of large $q_\mathrm{max}$ and $\psi$, high-$J$ attractors do not exist.
	
	\subsection{Total degree of alignment}
\begin{figure}
		\centering
		\includegraphics[width=\linewidth]{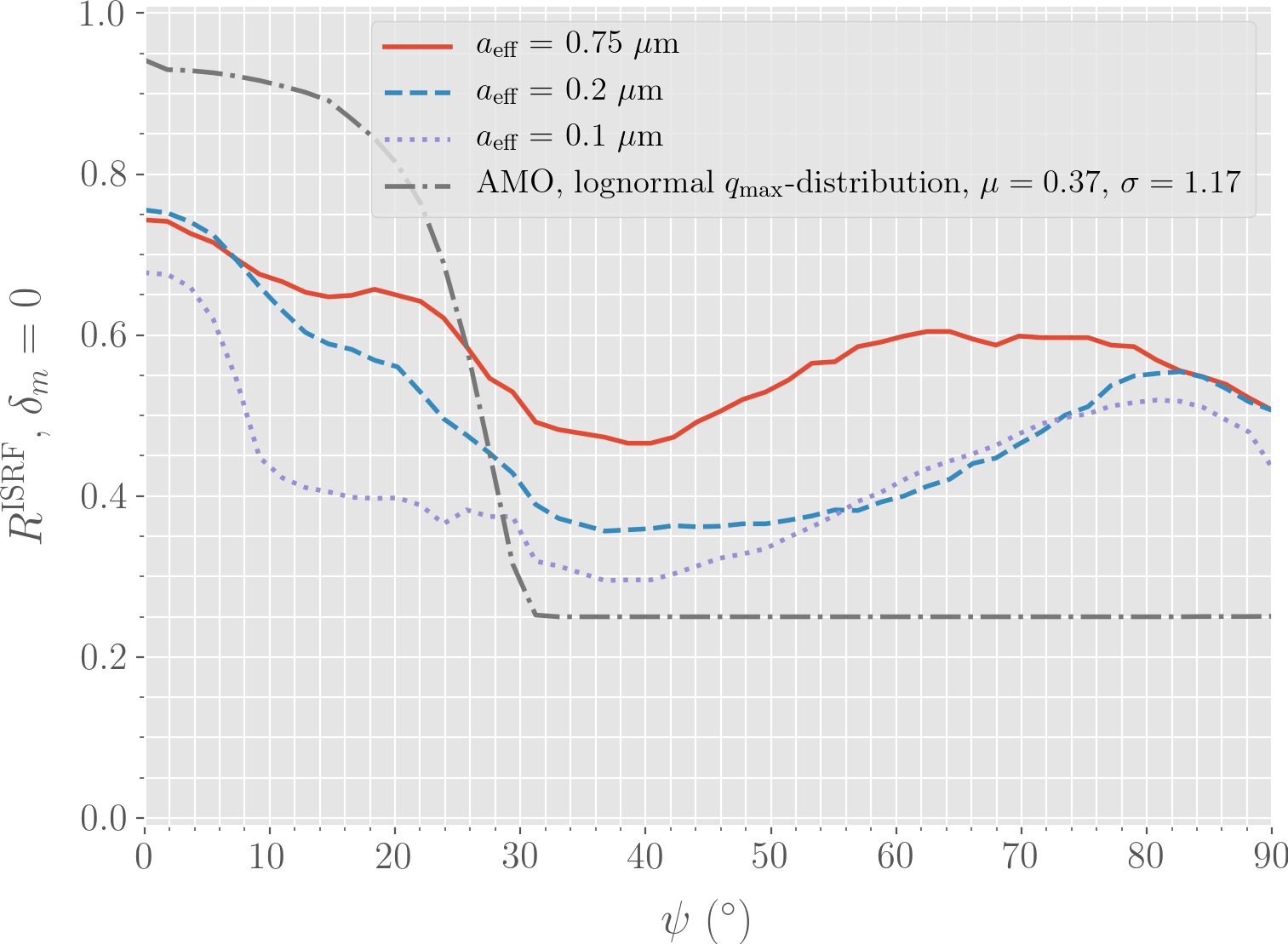}
		\caption{Total degree of alignment as a function of $\psi$ for GRE subensemble of different sizes, and for AMO with a lognormal $q_\mathrm{max}$-distribution.}
		\label{fig:figure11}
	\end{figure}

	The total degree of alignment is approximately given by Equation \ref{eq:totaldegree}, and depends on the degree of internally aligned grains with either low-$J$ or high-$J$ attractors. We recall, that for grains with a low-$J$ attractor only, the degree of internal alignment was chosen at $Q_{X,\mathrm{low}J}=0.25$. This results in total degrees of alignment presented in Figure \ref{fig:figure11}.
	
	The resulting degrees of alignment, where prolate and oblate subensembles are combined, resemble the $f_{\mathrm{high}J}$ results from Figure \ref{fig:figure5}, as expected. The choice of $Q_{X,\mathrm{low}J}$ guarantees that AMO at $\psi>30^\circ$ gives the lowest possible degree $R=0.25$. However, especially the large 0.75 {\textmu}m grains, have clearly larger degree of alignment.
	
	In the approximative scheme, where total degrees of alignment are derived from Equation \ref{eq:totaldegree}, $f_{\mathrm{high}J}$ provides straightforward means to solve for $R$ with any assumption on $Q_{X,\mathrm{low}J/\mathrm{high}J}$ values.

	\section{Discussion}\label{sec:discussion}
	
	\subsection{Main points of this study}
	
	LH07 identified grain helicity as the driver of the RAT alignment and presented a toy model of a helical grain that allowed a simple analytical description, i.e. the model of AMO. A significant progress in understanding of the properties of grain alignment was achieved by the AMO. Nevertheless, one would not expect the toy model to perfectly represent the RAT alignment of an ensemble of irregular grains. In fact, the deviations of the RAT functional form from AMO predictions were reported in LH07.\footnote{In LH07 modifications of AMO by changing the angle of the mirror to the grain axis of the maximal moment of inertia was considered as a way to reproduce torques acting for grains at different wavelength. In the absence of sufficient numerical input, the basic AMO with the mirror at angle of 45 degrees was used in the subsequent studies \citep[see][]{Hoang2008, Hoang2009, Hoang2016}.} However, with just a few shapes studied there it was not possible to evaluate the consequences of these deviations on the degree of alignment achievable by an ensemble of irregular grains. The present paper provides the statistical data needed by using an extensive number of irregular grains subjected to the interstellar radiation field. 
	
	The polarization predicted by the AMO was insufficient to explain the maximal values of polarization observed in the ISM, at least in the dust model where only silicate grains are aligned.\footnote{If the aligned grains are composite, i.e. contain both silicate and carbonaceous fragments, the constrains in the alignment efficiency are reduced. The abundunce of such composite grains is a controversial issue, however.} This made the AMO model with enhanced magnetic dissipation introduced in Lazarian \& Hoang (2008) relevant. The detailed predictions of AMO were presented in \citet[HL16]{Hoang2016}. 
	
	An important point of the present study is that our results show higher alignment efficiency compared to AMO, opening a possibility that the observed grain alignment can be accounted for by the RAT alignment of ordinary paramagnetic grains. However, magnetic inclusions cannot be immediately disregarded based on our results, where $R\sim0.4\textendash0.5$ for grain sizes $<0.3$ {\textmu}m. Recently, an analysis of spectral polarization by \citet{Draine2020} required $R\sim$ 0.7 to explain the 10 {\textmu}m polarization feature, assuming a reasonable axis ratio of spheroidal grains (between 1/3 and 3, favoring more modest ratios). The axis ratios of the undeformed spheroids in this study were 0.5 and 2. However, the higher alignment degrees of irregular grains, compared to AMO, allow the possibility that in some cases, RAT alignment degrees of irregular grains may be sufficient to explain some observations even if grains do not have magnetic inclusions.
	
	HL16 presented a contour map for the critical magnetic relaxation parameter, $\delta_{m,\mathrm{cri}}$ for AMO (adapted in Figure \ref{fig:figure1}), required to produce high-$J$ attractor points. There, it was found that for $\delta_{m}>10$, high-$J$ attractors are universal, and for smaller values, there are regions of interest both in terms of $q_\mathrm{max}$ and alignment angle $\psi$. 
	
	As a rule of thumb, $q_\mathrm{max}$ values near unity are most problematic for the existence of high-$J$ attractors. HL16 found that when approaching $\psi = 45^\circ$ from smaller angles, $\delta_{m,\mathrm{cri}}$ is smallest for grains with $q_\mathrm{max}$ less than unity. The same is true for $q_\mathrm{max}$ larger than unity when approaching from the large angle side. This implies that $f_{\mathrm{high}J}$ values should drop for numerical ensembles, given that the distribution of $q_\mathrm{max}$-factor contains values near unity, which was systematically observed in section 3. The exact value for the $\delta_{m,\mathrm{cri}}$ and the critical value for the $q_\mathrm{max}$ likely differ from those obtained from AMO, however, the trends suggested by AMO nevertheless appear.
	
	In our wavelength-dependent study, we find that as grains become smaller, $q_\mathrm{max}$ generally tends to gradually fall, although the special significant drop around $\lambda / a_{\mathrm{eff}} = 5$ introduces another general rule to the shape of $q_{\mathrm{max}}$ as a function of $\lambda / a_{\mathrm{eff}}$. Thus, the decrease in $f_{\mathrm{high}J}$ as a function of $\psi$ (Figures \ref{fig:figure6}–\ref{fig:figure7}, 0.1 {\textmu}m and 0.2 {\textmu}m cases) should slightly drift towards smaller values of $\psi$ at this size range. The effect is produced by our study, albeit the drift is very subtle, and is in good agreement with results predicted using AMO.
	
	In leftmost columns the 0.1 {\textmu}m and 0.2 {\textmu}m cases in Figures \ref{fig:figure6}–\ref{fig:figure7}, values of $f_{\mathrm{high}J}$ become more uniform as $\psi>50^\circ$. The size region is probed in more detail for 0.75 {\textmu}m grains ($N=400$). If $q_\mathrm{max}$ and $f_{\mathrm{high}J}$ are connected, the spread of $q_\mathrm{max}$ values as grains grow results in $f_{\mathrm{high}J}$ values that can be explained by AMO prediction from HL16. HL16 states that for $q_\mathrm{max}$ values greater than unity at $0^\circ<\psi<45^\circ$, grains with large $q_\mathrm{max}$ should be well aligned, and again, similarly for $\psi>45^\circ$ and small $q_\mathrm{max}$. Here, the idealized AMO predicts similar results, however for grains with large $q_\mathrm{max}$ the upper limit of the range of well aligned particles reduces to about $\psi=35^\circ$. As the spread of the $q_\mathrm{max}$ distribution is not extreme for large grains, the more uniform $f_{\mathrm{high}J}$ must be attributed also to RAT shape differences between AMO and irregular grains.
	
	Using numerical evidence of ensembles of irregular grains, it was found that AMO can successfully reproduce the basic features of the RAT alignment observed with the big sample of the GRE grains. In particular, it explains why the alignment takes place with long axes perpendicular to magnetic field, it explains that the alignment can happen at low-J and high-J attractor points and explains the properties of alignment achieved at these points. It also explains a decreasing maximum rotational rate with increasing $\psi$, which becomes rapid as $\psi$ exceeds 60$^\circ$. For irregular grains, such decrease is observed around $\psi\sim45^\circ$, beyond which $f_{\mathrm{high}J}$ fluctuates. The fluctuations in the grain-ensemble-averaged $f_{\mathrm{high}J}$ can likely be attributed to the irregular shape itself, due to which $\phi$-averaged RATs at certain angles provide attractor points. 
	
	The importance of $q_\mathrm{max}$ was further solidified by the evidence, as it correlates with the likelihood of existence of high-$J$ attractor points. The numerical results also imply that while the exact point of critical values for the parameters may differ for different types of grains due to complicated RAT shapes of irregular grains, the general behavior is produced correctly also by AMO. 
	
	{ At the same time, our results show the limitations of AMO in terms of qualitative description of the properties of some grains in the ensemble of shapes studied. To what extend this is limited to the shapes of higher elongation and whether the correspondence can be obtained with a modified AMO, e.g. by AMO with a mirror turned at different angle, as suggested as a modification in LH07, will be addressed elsewhere. We feel that the advantage of AMO is its extreme simplicity and having a set of different modifications might not be convenient. The development that we find very positive is that RATs acting on the studied ensemble of grains can provide a higher alignment degree compared to the classical model of AMO for large angles between the radiation direction and the magnetic field (i.e., $\psi>45^{\circ}$, see Figure \ref{fig:figure11}). This gives hope that the observed interstellar alignment can be explained without appealing to enhanced magnetic susceptibility of grains.}
	
	\subsection{Role of RATs and limitations of the present study}
	As we mentioned above, the most important question addressed in this paper is whether or not the enhanced magnetic dissipation is necessary for explaining high degrees of alignment. The LH07 theory was formulated for no additional relaxation, while it was shown in \citet{Lazarian2008c} that the higher degrees of alignment can be achieved if the high-$J$ stationary points are stabilized in the presence of strong magnetic response of the grains. The arguments in favor of the magnetic grains are presented in \citet{Hoang2016}, while in \citet{Lazarian2019}, the observational procedures for quantifying the grain magnetic properties were discussed. 
	
	Our present work opens up a discussion of interstellar grain composition. An important question is whether the presence of magnetic inclusions \citep{Lazarian2008c, Hoang2016} is necessary to explain the observed interstellar alignment. The increased grain alignment efficiency reported in this study makes it easier to explain interstellar grain alignment. The task gets simplified if a significant part of interstellar carbon is locked in the composite grains consisting of carbonaceous and silicate fragments. We note, however, that our ability to explain the interstellar polarization with paramagnetic grains does not actually mean that the grains should not contain magnetic inclusions. 
	
	For instance, most of the interstellar iron is locked in grain material but we do not know what chemical substance iron is a part of and what are the magnetic properties of the substance. Thus, it is important to test the composition of grains by measuring the polarization and its changes in special settings, for instance, near stars of transient radiation sources, e.g. novae. Furthermore, existing studies have constrained the possibility of magnetic inclusions by comparing predictions \citep{Draine2013} against observations \citep{Planck2020}. The corresponding tests of the magnetic properties of grains based on the grain alignment theory are discussed in \cite{Lazarian2019}, while in \citet{Lazarian2020b}, the tests related to the alignment-dependent grain disruption are proposed.
	
	In the present paper we considered the $B$-alignment, i.e. the alignment in respect to the magnetic field. As was demonstrated in LH07, the alignment can also happen in respect to the direction of radiation, provided that the precession induced by the $\Gamma_3$ component of the radiative torque is faster than the grain precession about the ambient magnetic field. The process of $k$-alignment can be analyzed with our present approach by assuming $\psi=0$.
	
	The RAT alignment process is a complex one with only a small fraction of grains getting into the high-$J$ attractor points initially. In the presence of time-dependent radiation sources, LH07 demonstrated that the transient alignment is to take place. During the process most grains are aligned in the low-$J$ attractor point on the time scale of the order of grain precession induced by the $\Gamma_3$ RAT component. However, on the time of several $\tau_{\mathrm{ran}}$ the grains are getting to the stationary alignment at high-$J$ attractor points if such points are present for the given combination of $q_\mathrm{max}$ and $\psi$.
	
	For our calculation in the paper we assumed that the grains are subject to the sufficiently strong radiation field that makes grain at high-$J$ attractor points rotate with velocities significantly faster than the thermal one. These conditions are well satisfied in the diffuse ISM for $\sim 0.1$ {\textmu}m grains. For grains shielded from the radiation in molecular clouds or accretion disks as well as for grains with sizes appreciably different from the typical wavelength of the impinging radiation, the rotational rates of grains at high-$J$ points can be reduced to become comparable to the thermal rotation speed. Such grains can be randomized by gaseous collisions and other randomizing factors. The RAT alignment of such grains in such conditions is decreased for $\psi$ larger than $80$ degrees. According to Figure 8 in \citet{Lazarian2019}, RAT strength falls off for that range. 
	
	The current study is for the interstellar radiation field. The ISM is the area where the grain alignment has been mostly studied \citep[see][]{Andersson2015}. However, the RAT alignment of dust is a ubiquitous process that can take place in different astrophysical environments and in planet atmospheres \citep[see][]{Lazarian2007}. Therefore, detailed studies of the peculiarities of grain alignment for different spectra of radiation illuminating ensembles of irregular grains of different compositions are required. Therefore this work can be viewed as the first step in this direction.  
	
	\subsection{Application to carbonaceous grains}
	The properties of RAT torques we explored in this paper are applicable for grains made of different material. The change of the optical constants is not expected to radically change the grain alignment. Therefore, on the basis of the obtained results we can consider the alignment of not only silicate, but also carbonaceous grains. The RAT alignment of carbonaceous grains was considered in \citet{Lazarian2020} and it was shown that carbonaceous grains can exhibit a different mode of RAT alignment, i.e. the alignment with long grain axes parallel to magnetic field. The anomalous mode arises due to charged carbonaceous grain precession along the electric field. The latter arises due to the grain motion in respect to magnetic field. 
	
	The difference between the alignment of carbonaceous and silicate grains considered in this study arises due to the difference of the magnetic moment of the grains. The latter is significantly smaller for carbonaceous grains. Our qualitative conclusions on the modification of grain alignment compared to the AMO prediction stay true, but in terms of observed polarization, the observed degrees can be lower due to grain rotation about the magnetic field. 
	
	If carbonaceous grains get into the state of high-$J$ rotation, their magnetic moment increases and it was shown in \citet{Lazarian2020} that such grains can get aligned with long grain axes perpendicular to the magnetic field, i.e., to be aligned similar to silicate grains. Our present study shows that the percentage of grains in the high-$J$ rotation state for irregular grains is larger than the AMO prediction. Therefore, we may expect more carbonaceous grains to get aligned in the regular fashion, i.e. with long grain axes perpendicular to the magnetic field.  
	
	The alignment of grains with long axes parallel to the magnetic field can, as explained in \citet{Lazarian2020} result in a more complicated pattern of polarization that is more difficult to interpret in terms of the underlying magnetic field. At the same time, if actual interstellar grains are composite, i.e. contain both carbonaceous and silicate fragments, their alignment is similar to that of silicate grains. 	
	
	\subsection{Are highly irregular grains shapes present in the ISM?}
	{ The ensemble of irregular grain shapes used for our study and comparison to AMO are generated using Gaussian random algorithm, which are not subject to any astrophysical constraints. Therefore, a considerable fraction of generated grains has extreme shapes characterized by the large ratio of RAT components, $q^\mathrm{max}<0.5$ or $q^\mathrm{max}>5$, which induces a higher degree of alignment for $\psi>45^{\circ}$ than predicted by AMO. However, whether such highly irregular shapes can exist in the ISM remain unclear. Highly irregular shapes are likely to have a low tensile strength, so that they are more easily disrupted by by RATs (\citealt{Hoang2019}; \citealt{{Hoang:2019},{2020Galax...8...52H}}). More observational studies on grain shapes are required to test RAT alignment and disruption theory as well as the relation of the alignment and the disruption (see \citealt{Lazarian2020b}).}
	
	\section{Summary}\label{sec:summary}
	In the paper above we have explored the grain alignment for a collection of irregular grains of different shapes. We compared the properties of the RAT alignment with the standard AMO. While the deviations in the RAT functional shape from the AMO prediction are not significant, we report a general improvement of alignment compared to the AMO expectations. On thi basis we may argue that the explanation of the observed alignment can { potentially} be achieved with ordinary paramagnetic grains without magnetic inclusions. Our results can be briefly summarized as follows:
	\begin{itemize}
		\item Better alignment of irregular grain RATs compared to the AMO predictions is found, especially when alignment at large $\psi$ is considered.  In particular, $\sim$40\% of irregular grains at $\psi=80^\circ$, deviate from the AMO predictions and exhibit high-$J$ attractors. In this situation, whether magnetic inclusions are necessary to explain the properties of interstellar polarization requires more theoretical and observational research.
   \end{itemize}

    \section{Notes added in proof}
    Our more recent study shows that, indeed, as we mentioned in the paper, the modification of AMO via changing the angle that the mirror makes with the circumference of the grain improves the correspondence of the AMO predictions with the results obtained for the ensemble of grains studied numerically in this paper. In other words, while the original AMO with a mirror attached at $\alpha=45$ degrees provides a good correspondence for many grain shapes, the variations of $\alpha$ allows to fit the properties of very different grain shapes interacting with electromagnetic radiation at very different wavelength in agreement with the original study in \citet{Lazarian2007b}. 

    The combination of the standard AMO ($\alpha=45^\circ$) and a modified AMO with the $\alpha \lesssim 30^\circ$ would decrease the fraction of high-J attractors at $\psi<45^\circ$ and increase the fraction of high-J at $\psi>45^\circ$. From Fig B16 in LH07 one can see that $q^\mathrm{max}$ decreases with decreasing $\alpha$ so we can combine AMO with different angles such, e.g  $f_{45}\mathrm{AMO}(\alpha=45) + (1-f_{45})\mathrm{AMO}(\alpha=20)$ with $f_{45}$ is the fractions of the grains with the mirror at $45^\circ$. Our estimates show that with $f_{45} ~ 0.7$, we can bring AMO to be in agreement with GRE as shown in Figure 5. Naturally, this is just a fit, but it shows the power of AMO in terms of explaining properties of RATs for ensembles of irregular grains.

    The important new finding in this paper is those adjustments increase the parameter space for which one expects perfect RAT alignment. This indicates that the necessity of superparamagnetic enhancement of the RAT alignment \citep{Lazarian2008c, Hoang2016} may not be necessary in order to explain observations. The latter point requires further studies.
	\acknowledgments
	J.H. thanks the UW Department of Astronomy for its hospitality during his visit. A.L. acknowledges the support of the NSF grant AST 1715754, and 1816234. Flatiron Institute is supported by Simons Foundation. T.H. acknowledges the support by the National Research Foundation of Korea (NRF) grants funded by the Korea government (MSIT) through the Mid-career Research Program (2019R1A2C1087045). { We thank the referee for numerous comments which helped us to improve the presentation of our results.}

\end{document}